\documentclass[11pt]{article}

\usepackage[preprint]{acl}

\usepackage{times}
\usepackage{latexsym}

\usepackage[T1]{fontenc}

\usepackage[utf8]{inputenc}

\usepackage{microtype}

\usepackage{inconsolata}

\usepackage{graphicx}
\usepackage{subcaption}
\usepackage{amsmath}
\usepackage{amsfonts}
\usepackage{enumitem}
\usepackage{tcolorbox}
\usepackage{multirow}
\usepackage{xcolor}
\usepackage{colortbl}
\usepackage{array}
\usepackage{makecell}
\usepackage{pifont}

\usepackage{tcolorbox}
\tcbuselibrary{listings,skins,breakable}  

\usepackage{listings}
\lstdefinelanguage{Markdown}{
  alsoletter={\#\*\-\+\_},
  morecomment=[l]{<!--},
  sensitive=false
}

\lstdefinestyle{markdown-style}{
  language=Markdown,
  backgroundcolor=\color{green!2},
  basicstyle=\ttfamily\scriptsize,
  keepspaces=true,
  keywordstyle=\color{blue}\bfseries,
  commentstyle=\color{gray},
  stringstyle=\color{red},
  breaklines=true,
  breakindent=0pt,
  postbreak=,
  showstringspaces=false,
  columns=fullflexible,
}

\newtcblisting{markdowncode}{
  breakable,                  
  enhanced,                   
  colback=green!2,            
  colframe=blue!70!black,     
  boxrule=1pt,                
  arc=10pt,                   
  sharp corners=east,         
  leftrule=0pt,               
  toprule=0pt, bottomrule=0pt, rightrule=0pt, 
  fontlower=\small,           
  listing only,               
  listing options={style=markdown-style},
}

\lstdefinestyle{artifact-style}{
  language=Markdown,
  backgroundcolor=\color{yellow!5},
  basicstyle=\ttfamily\scriptsize,
  keepspaces=true,
  keywordstyle=\color{blue}\bfseries,
  commentstyle=\color{gray},
  stringstyle=\color{red},
  breaklines=true,
  breakindent=0pt,
  postbreak=,
  showstringspaces=false,
  columns=fullflexible,
}

\newtcblisting{artifact}{
  breakable,                  
  enhanced,                   
  colback=yellow!5,            
  colframe=blue!70!black,     
  boxrule=1pt,                
  arc=10pt,                   
  sharp corners=east,         
  leftrule=0pt,               
  toprule=0pt, bottomrule=0pt, rightrule=0pt, 
  fontlower=\small,           
  listing only,               
  listing options={style=artifact-style},
}
\newcommand{\ourtool}{RAFT}
\newcommand{\strategy}{Adaptive Purification-Aggregation}
\newcommand{\strategymm}{Purification}
\newcommand{\strategyrr}{Aggregation}

\title{Explicating Tacit Regulatory Knowledge from LLMs to Auto-Formalize Requirements for Compliance Test Case Generation}

\author{\And
  Zhiyi Xue$^1$, Xiaohong Chen$^1$, Min Zhang$^2$\\
  $^1$Shanghai Key Laboratory of Trustworthy Computing, ECNU, Shanghai, China \\
  $^2$Dishui Lake International Software Engineering Institute, ECNU, Shanghai, China \\
  \texttt{52275902017@stu.ecnu.edu.cn, \{xhchen,zhangmin\}@sei.ecnu.edu.cn} \\\And
}

\begin{document}
\maketitle
\begin{abstract}


Compliance testing in highly regulated domains is crucial but largely manual, requiring domain experts to translate complex regulations into executable test cases. While large language models (LLMs) show promise for automation, their susceptibility to hallucinations limits reliable application. Existing hybrid approaches mitigate this issue by constraining LLMs with formal models, but still rely on costly manual modeling. To solve this problem, this paper proposes \ourtool, a framework for requirements auto-formalization and compliance test generation via explicating tacit regulatory knowledge from multiple LLMs. \ourtool\ employs an \strategy\ strategy to explicate tacit regulatory knowledge from multiple LLMs and integrate it into three artifacts: a domain meta-model, a formal requirements representation, and testability constraints. These artifacts are then dynamically injected into prompts to guide high-precision requirement formalization and automated test generation. Experiments across financial, automotive, and power domains show that \ourtool\ achieves expert-level performance, substantially outperforms state-of-the-art (SOTA) methods while reducing overall generation and review time.

\end{abstract}

\section{Introduction}
\label{sec:introduction}

In highly regulated domains such as finance, automotive, and energy, software compliance testing is crucial to ensure adherence to regulations, industry standards, and internal policies \citep{badrzadeh2014challenges,tabani2019assessing,xue2024llm4fin,zhang2025comprehensive}. Failures in compliance can lead to severe safety incidents, regulatory penalties, and substantial financial losses \citep{gm_ignition_switch_recalls_wikipedia,sivakumar2024came}. 
Despite its importance, compliance testing in practice remains predominantly manual. Domain experts are required to interpret lengthy and complex regulatory texts and translate them into executable test cases. This process is labor-intensive and error-prone, making its automation a key focus of industry development \citep{castellanos2022compliance,jain2025complexity}. 

Existing automated solutions face a trade-off between modeling cost and generation reliability. Traditional model-based approaches rely on explicitly constructed models, graphs, or formal representations for test generation~\citep{wang2020automatic,stefani2025automated,zyberaj2025genai,yang2025llmcfg}. While precise and controllable, these approaches require substantial upfront expert effort, limiting their scalability to other regulatory domains. In contrast, LLM-based methods reduce manual effort by generating tests directly from regulatory texts or via multi-agent workflows~\citep{castellanos2022compliance,boukhlif2024towards,korraprolu2025test,hasan2025automatic,masuda2025generating}. However, their probabilistic nature leads to hallucinations and low interpretability, hindering their adoption in regulation-critical settings.

To reconcile this tension, hybrid test generation frameworks have combined model-driven techniques with LLM-based generation~\citep{xue2024llm4fin,chen2024automated,sun2025compliance,liu2025llm,shrestha2025less}. In these approaches, LLMs translate natural language regulations into predefined formal requirements, which constrain generation and mitigate hallucinations. However, the representation of these requirements still requires substantial manual definition and maintenance by domain experts; when the domain shifts (e.g., from finance to automation), they must be re-engineered, resulting in high maintenance costs.

Automating requirements formalization itself is the most direct way to break this deadlock, but it demands regulatory knowledge on demand. Hence, the paradigm must move from ``manual modeling'' to ``automatic knowledge explication''. Modern LLMs, having undergone large-scale pre-training, already encode a wealth of tacit regulatory knowledge, yet it remains implicit and unstructured. Only by unlocking and formalizing this knowledge into explicit, verifiable artifacts can we enable test-case generation that is both reliable and scalable.


In this paper, we propose \ourtool, a novel framework for Requirement Auto-Formalization and compliance Test generation via explicating tacit regulatory knowledge from multiple LLMs.
\ourtool\ introduces a dedicated \textit{Regulatory Knowledge Explication} phase, in which an \strategy\ strategy is employed to extract and formalize three knowledge artifacts from LLMs: a domain meta-model, a formal requirements representation, and testability constraints. Together, these artifacts establish a coherent conceptual and logical foundation prior to processing individual regulatory texts.
Building upon these knowledge artifacts, \ourtool\ performs knowledge-injected prompting during the \textit{Prompt and Test Case Generation} phase. The extracted artifacts are dynamically incorporated into prompts as structural and semantic constraints, guiding the LLM to formalize regulations into precise, interpretable requirements that can be transformed into high-quality test cases.

\ourtool\ promotes LLMs from ``Text Translators'' to ``Knowledge Architects''. 
By explicitly extracting and injecting tacit regulatory knowledge, the framework dramatically boosts the controllability, interpretability, and reliability of LLM-driven requirements formalization and compliance testing, delivering a fully automated and scalable solution.

We conducted a comprehensive evaluation on \ourtool. In the financial domain, it achieved expert-level performance with an average 91.7\% F1 and 86.5\% business scenario coverage, outperforming SOTA methods by up to 34.7\%, while reducing the total generation and review time from 5.7 hours or even 2 weeks to just 2.5 hours. Ablation studies confirm that each explicated knowledge artifact, as well as the proposed \strategy\ strategy, is essential for high-quality test generation. Furthermore, \ourtool\ demonstrates strong cross-domain generalizability, achieving average F1 scores of 92.7\% and 87.2\% in the automotive and power domains, respectively. Our code and data are available at \url{https://github.com/1767675261/RAFT}.

Our contributions are summarized as follows:
\begin{enumerate}[leftmargin=*,align=left,label=(\arabic*), itemsep=0pt, parsep=0pt, nosep]
    \item \textbf{Knowledge Enhanced Requirements formalization and Test Generation:} 
    We propose \ourtool, a framework that automatically explicates regulatory knowledge from LLMs and injects it into the test generation, shifting compliance testing from expert-driven modeling to knowledge-driven automation.

    \item \textbf{Regulatory Knowledge Explication:} 
We define and extract three reusable, explicit artifacts, including domain meta-model, formal requirements representation, and testability constraints, turning originally implicit and fragmented tacit regulatory knowledge into verifiable and cross-domain portable assets.  
    
    \item \textbf{Knowledge-Constrained Prompting:} 
   We devise a prompting scheme that dynamically embeds the three artifacts as structured constraints in the prompt, enabling LLMs to produce precise, regulation-compliant outputs without retraining, balancing accuracy and generality.
\end{enumerate}

\section{Related Work}
\label{sec:related_work}

\begin{figure*}
    \centering
    \includegraphics[width=0.9\linewidth]{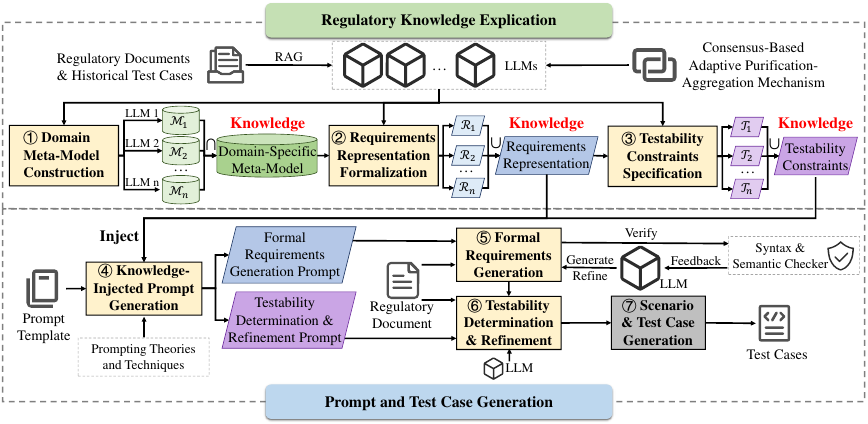}
    \caption{The regulatory knowledge explication and injection-driven framework for fully automated test generation.}
    \label{fig:framework}
\end{figure*}

This section reviews prior work on automated test case generation, focusing on how regulatory knowledge is represented and utilized.

\noindent
\textbf{Model-Based Test Generation.}
Model-based testing introduces explicit intermediate graphs or formal models for testing. Representative examples include control-flow graphs for path coverage \citep{yang2025llmcfg}, System Graphs converted into executable tests \citep{zyberaj2025genai}, and formal ODD or rule models for safety-critical systems \citep{stefani2025automated,xue2024llm4fin}.
While controllable, these methods depend heavily on manually constructed models that are complex and labor-intensive, limiting scalability \citep{xue2024domain}.

\noindent
\textbf{LLM-Based Test Generation.}
Recent studies leverage LLMs or multi-agent systems to generate test cases. Prompt-based methods produce test inputs or structured test steps \citep{wang2025testeval,korraprolu2025test}, while multi-agent pipelines emulate human testing workflows \citep{milchevski2025multi} or extract testing strategies prior to generation \citep{masuda2025generating}. Domain-specific fine-tuning has also been explored \citep{boukhlif2024towards,hasan2025automatic}. 
While these approaches reduce manual effort, their reliance on probabilistic outputs leads to hallucinations and instability, limiting their applicability in highly regulated domains \citep{zhang2024testbench,korraprolu2025test}.

\noindent
\textbf{Hybrid Frameworks.}
Hybrid approaches combine formal representations with LLM-based generation to balance controllability and flexibility. Examples include integrating LLMs with structured test scenarios in financial software \citep{xue2024llm4fin,liu2025llm}, combining ontologies and learning-based methods for BIM compliance checking \citep{chen2024automated}, and guiding LLMs with retrieval-augmented generation or predefined requirement templates \citep{sun2025compliance,shrestha2025less}.
Although these approaches mitigate hallucinations by constraining the generation space, they still rely on hand-crafted modeling, leaving regulatory knowledge statically engineered and thus limiting scalability and adaptability.

\section{The RAFT Approach}
\label{sec:method}

\subsection{Regulatory Knowledge and Overview}


To achieve fully automated test case generation based on a hybrid framework, it is necessary to automate the conversion of regulatory documents into formal requirements. Accomplishing this goal requires satisfying two prerequisites: first, clarifying how requirements should be expressed; second, confirming that the requirements are testable. 

Regarding the question of ``how to represent requirements'', we focus on two aspects: the composition of requirement elements and the form of their expression. Accordingly, the representation process is divided into two stages: first, extracting key requirement elements to construct a requirement metamodel; and then completing the formal requirements representation based on this model. In parallel, we define testability constraints to identify testable requirements. Ultimately, we derive three core types of regulatory knowledge: the domain meta-model $\mathcal{M}$, the requirements representation $\mathcal{R}$, and the testability constraints $\mathcal{T}$.

After the regulatory knowledge is formed, it can be systematically embedded into a prompt-based framework to guide LLMs in two key tasks: converting natural-language clauses into formal requirements and automatically generating executable test cases. To this end, we propose \ourtool\ (Figure \ref{fig:framework}), a two-phase approach that ensures semantic consistency and testability by combining knowledge explicitation with prompt engineering:

\textbf{Phase I: Regulatory Knowledge Explicitation.} The domain meta-model $\mathcal{M}$, the requirements representation $\mathcal{R}$, and the testability constraints $\mathcal{T}$ from regulatory documents and existing test cases to establish a structured knowledge system.

\textbf{Phase II: Prompt and Test Case Generation.} Inject the explicated knowledge into prompt templates to steer the LLMs through sequential requirement formalization and test case generation, resulting in an end-to-end automated workflow.


\subsection{Regulatory Knowledge Explication}
\label{sec:know_ext}



\noindent
\textbf{Challenges in explicating regulatory knowledge from LLMs.}
The core objective of this phase is to obtain the three knowledge assets: the domain meta-model $\mathcal{M}$, 
the requirements representation $\mathcal{R}$, and the testability constraints $\mathcal{T}$. 
We decompose the process into three extraction steps, all implemented using LLMs. To assess the feasibility of automated regulatory knowledge explication, we conducted preliminary experiments using SOTA LLMs to construct these assets with direct instructions. The results revealed two recurring challenges:
\begin{enumerate}[leftmargin=*,align=left,label=(\arabic*), itemsep=0pt, parsep=0pt, nosep]
    \item \textbf{Granularity and Structural Inconsistency.} LLMs often fail to identify an appropriate level of abstraction, leading to missing elements or inconsistent structural representations.
    \item \textbf{Hallucination and Output Variability.} Generated structures may be ungrounded or unstable across runs, hindering the construction of reliable regulatory knowledge.
\end{enumerate}

To address these issues, we propose two corresponding solutions: a Chain-of-Thought (CoT) \citep{wei2022chain} enhanced with Retrieval Augmented Generation (RAG) \citep{lewis2020retrieval} prompting method, and a multi-LLM-based \strategy\ strategy.

\vspace{2mm}
\noindent
\textbf{CoT \& RAG Prompting.}
This mechanism constrains LLM reasoning along four dimensions: (1) \textit{Task Orientation}, framing the task as explicit reconstruction of tacit regulatory knowledge; (2) \textit{Hierarchical Decomposition}, enforcing a structured whole-to-part reasoning process tailored to each knowledge type; (3) \textit{Granularity Constraint}, restricting the level of detail to what is necessary for the target artifact; and (4) \textit{Retrieval Grounding}, incorporating regulatory documents and historical test cases to reduce hallucinations. 
This prompt framework is applied to \textit{Steps} \ding{172}, \ding{173}, and \ding{174}, with task-specific adaptations. 

\vspace{2mm}
\noindent
\textbf{\strategy.}
The key idea is to apply reconciliation mechanisms tailored to different knowledge types, reflecting their distinct structural and semantic characteristics. Specifically, 
In \textit{Step} \ding{172}, the domain meta-model requires both structural integrity and parsimony \citep{evans2004domain}. We therefore introduce a \textit{$K$-\strategymm} operation, which treats individual LLM outputs as noisy observations and extracts a stable meta-model core via majority consensus:



\begin{tcolorbox}[colback=green!5!white, colframe=green!30!black, title={Definition: K-\strategymm}, fonttitle=\bfseries\small, fontupper=\small, rounded corners]
\vspace{-2mm}
Let $\mathbf{M} = \{\mathcal{M}_1, \mathcal{M}_2, \ldots, \mathcal{M}_N\}$ be the meta-models generated by $N$ independent LLMs. Given a consensus threshold $K$ ($1 < K \leq N$), the final meta-model $\mathcal{M}$ is defined as:
\vspace{-3mm}
\begin{equation}
    \mathcal{M} = \{ e \in \bigcup_{i=1}^N \mathcal{M}_i \mid \sum_{i=1}^N \mathbb{I}(e \in \mathcal{M}_i) \geq K \}
\end{equation}

\vspace{-3mm}

where $e$ denotes meta-model elements and $\mathbb{I}(\cdot)$ is the indicator function.
\vspace{-1mm}
\end{tcolorbox}


We set $N=3$ and $K=2$ in our implementation, retaining majority-supported elements to suppress stochastic hallucinations from individual LLMs, while grouping low-frequency elements as ``Others'' to preserve coverage of regulatory semantics.

In \textit{Steps} \ding{173} and \ding{174}, the primary risk shifts from structural noise to information omission \citep{chomsky2002syntactic}. Since missing regulatory logic or testability constraints can undermine downstream correctness, 
we adopt an \textit{\strategyrr} strategy that prioritizes logical completeness over frequency:

\begin{tcolorbox}[colback=blue!5!white, colframe=blue!30!black, title={Definition: \strategyrr}, fonttitle=\bfseries\small, fontupper=\small, rounded corners]
\vspace{-2mm}
Given formal requirements representations $\{\mathcal{R}_i\}_{i=1}^N$ and testability constraints $\{\mathcal{T}_j\}_{j=1}^N$ from $N$ LLMs, aggregation is performed via:
\vspace{-3mm}
\begin{equation}
    \mathcal{R} = \bigcup_{i=1}^N\mathcal{R}_i, \quad
    \mathcal{T} = \bigcup_{j=1}^N\mathcal{T}_j 
\end{equation}
\vspace{-6mm}
\end{tcolorbox}

With $N=3$, this strategy minimizes omission risk by aggregating all inferred regulatory logic and constraints. Overall, the \strategy\ strategy mitigates hallucination and divergence in single LLMs by enforcing structural rigor in meta-model construction and logical completeness in requirements representation formalization and testability constraints specification.

\begin{figure}
    \centering
    \includegraphics[width=\linewidth]{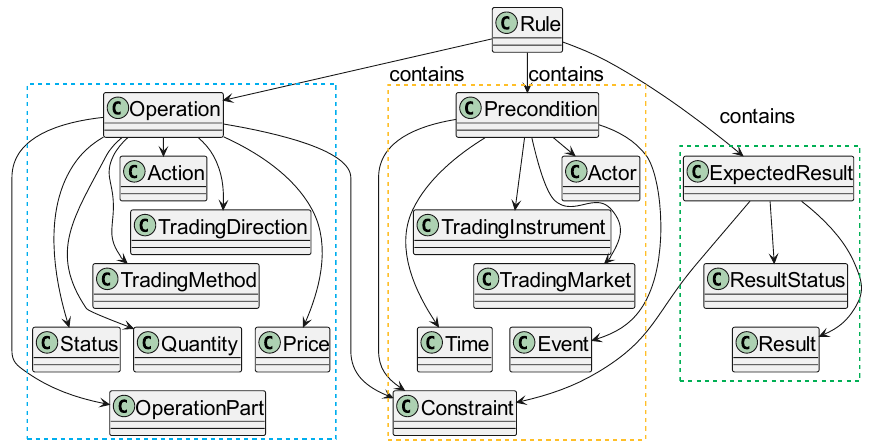}
    \caption{Meta-model for the finance domain.}
    \label{fig:metamodel_real}
\end{figure}

\vspace{2mm}
\noindent
\textbf{Example of Explicated Regulatory Knowledge.}
We illustrate the explicated regulatory knowledge using the finance domain, employing GPT-5~\citep{gpt}, Grok-4~\citep{grok}, and DeepSeek-R1~\citep{guo2025deepseek}. The financial meta-model $\mathcal{M}_f$ is expressed as PlantUML, consisting of 14 core elements (e.g., \textit{TradingVariety}, \textit{Price}) organized into a three-layer hierarchy in PlantUML \citep{plantuml}, as shown in Figure~\ref{fig:metamodel_real}. By explicitly separating \textit{Preconditions}, \textit{Operations}, and \textit{Expected Results}, the meta-model provides a verifiable causal structure and consolidates vague regulatory concepts into representative domain entities.

\begin{figure}
    \centering
    \includegraphics[width=\linewidth]{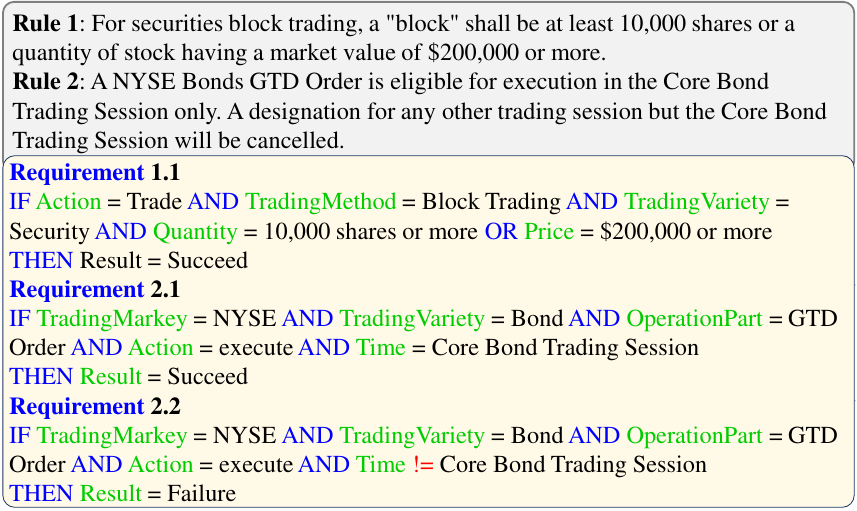}
    \caption{An example of 2 regulatory rules from NYSE \citep{nyse} and their requirement representation.}
    \label{fig:rule_example}
\end{figure}

Grounded in $\mathcal{M}_f$, the finance requirements representation $\mathcal{R}_f$ defines a library of domain symbols and a BNF-based \citep{backus1960report} formal syntax. As exemplified in Figure~\ref{fig:rule_example}, requirements follow a structured \textit{IF–THEN} logic composed of \textit{KEY–OP–VALUE} clauses connected by logical operators, where \textit{KEY} is strictly aligned with entities in $\mathcal{M}_f$ to ensure semantic consistency, \textit{VALUE} specifies the corresponding instance, and \textit{OP} denotes the operator that defines their relations.

Based on $\mathcal{R}_f$, finance testability constraints $\mathcal{T}_f$ are specified as OCL constraints \citep{ocl2014object} covering five aspects, including \textit{Structural Completeness}, \textit{Element Determinism}, \textit{Action Executability}, \textit{Result Observability}, and \textit{Rule Non-Conflict}. For example, \textit{Rule 1} in Figure~\ref{fig:rule_example} satisfies all constraints and is therefore testable, whereas \textit{Rule 2} is deemed untestable due to the non-deterministic element ``core bond trading session''.



\subsection{Prompt and Test Case Generation}
\label{sec:test_gen}


This phase focuses on transforming regulatory rules into executable test cases through formalizing requirements. There are four steps: Prompt Generation, Requirements Formalization, Testability Determination \& Refinement, and finally Test Case Generation. The prompts used and formalized regulatory knowledge artifacts in this section are provided in Appendices \ref{sec:app_prompt_ours} and \ref{sec:app_knowledge}, respectively.

\begin{figure}
    \centering
    \includegraphics[width=\linewidth]{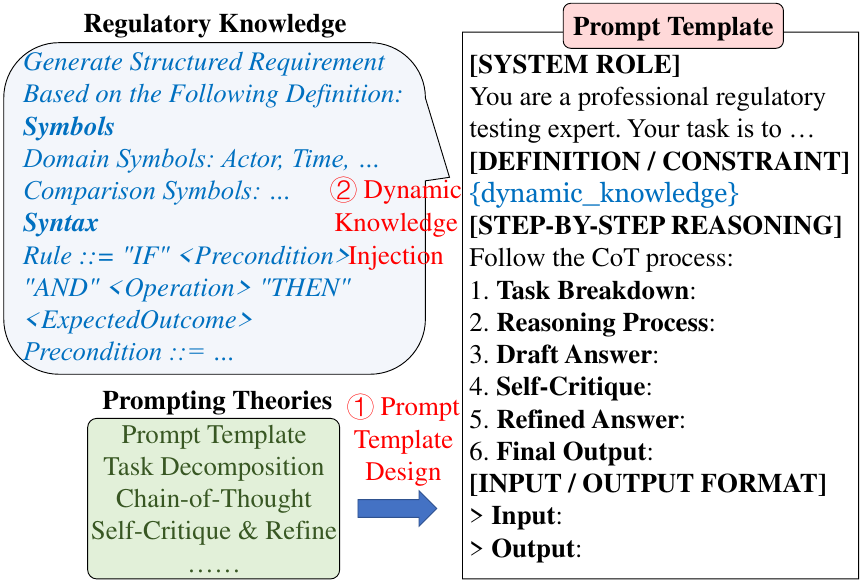}
    \caption{Dynamic knowledge injection-based prompt generation for Formal Requirement Generation.}
    \label{fig:know_inj}
\end{figure}

\noindent
\textbf{Knowledge-Injected Prompt Generation.}
In \textit{Step} \ding{175}, we design two specialized prompts that dynamically inject the explicated knowledge artifacts into the generation process, as shown in Figure \ref{fig:know_inj}:

\begin{enumerate}[leftmargin=*,align=left, itemsep=0pt, parsep=0pt, nosep]
    \item[(1)] \textbf{Formal Requirement Generation Prompt ($P_\mathcal{R}$).} This prompt embeds the domain symbol library and the requirements representation syntax from $\mathcal{R}$ into a zero-shot template, constraining the LLM to generate formal \textit{IF–THEN} requirements. By enforcing these grammatical constraints, all generated symbols and relations are strictly grounded in $\mathcal{R}$.
    \item[(2)] \textbf{Testability Determination \& Refinement Prompt ($P_\mathcal{T}$).} This prompt incorporates constraints from $\mathcal{T}$ as fine-grained evaluation criteria. Each testability constraint is assessed independently across all requirements, enabling the detection of subtle logical gaps that holistic judgments may miss due to long context.
\end{enumerate}


\vspace{2mm}
\noindent
\textbf{Formal Requirement Generation and Testability Determination \& Refinement.}
In \textit{Step} \ding{176}, $P_\mathcal{R}$ is used to guide an LLM to generate formal requirements from regulatory documents within a \textit{Verification-Feedback-Refinement} loop. The BNF syntax of $\mathcal{R}$ is first converted by an LLM into Xtext-compliant grammar, which Xtext then uses to automatically construct a checker. All generated requirements are iteratively validated against this checker and corrected by the LLM.

In \textit{Step} \ding{177}, the generated formal requirements are assessed against all testability constraints by instructing an LLM with $P_\mathcal{T}$. Violations trigger targeted refinement attempts by the LLM, including completing missing elements, concretizing abstract expressions, and so on. Requirements that remain logically inconsistent, unrefinable, or purely descriptive are excluded for expert review.

We evaluated this process on 19 financial regulatory documents using GPT-5, Grok-4, and DeepSeek-R1, with expert feedback establishing ground truth. The performance of our approach was high, achieving token-wise and word-wise Precision/Recall/F1 of 91.6\%/87.7\%/89.0\% and 89.8\%/85.3\%/86.8\%, respectively. Testability assessment showed that 77.6\% of requirements are automatable, with fewer than 5\% requiring manual intervention, indicating that it can effectively identify testable rules and minimize human effort.

\vspace{2mm}
\noindent
\textbf{Scenario and Test Case Generation.}
In \textit{Step} \ding{178}, we integrate testable formal requirements into a hybrid test generation framework, LLM4Fin~\cite{xue2024llm4fin}, which is most compatible with our requirements representation, to perform the final Scenario and Test Case Generation.

Test scenarios are built by identifying inter-rule dependencies, mapping behavioral elements (\textit{Action}, \textit{Status}) to activity diagrams and state machines to form end-to-end business flows. For each scenario, concrete test cases are generated using strategies such as Equivalence Partitioning and Boundary Value Analysis \citep{graham2006foundations}. Numerical inputs produce positive, negative, and boundary values, while finite non-numerical inputs are exhaustively enumerated, 
yielding a compact yet comprehensive set of test cases. 

\section{Experimental Evaluation}
\label{sec:experiment}

To evaluate the performance of \ourtool, we conducted a comprehensive experimental study to address the following research questions:
\begin{enumerate}[leftmargin=*,align=left, itemsep=0pt, parsep=0pt, nosep]
    \item[\textbf{RQ1:}] How does our approach perform compared with SOTA baselines in terms of the effectiveness and efficiency of generated test cases?
    \item[\textbf{RQ2:}] What is the contribution of the proposed \strategy\ strategy and each explicated regulatory knowledge artifact to the quality of generated test cases?
    \item[\textbf{RQ3:}] How well does \ourtool\ generalize across different regulated domains?
\end{enumerate}

\subsection{Experiment I: Effectiveness and Efficiency of Test Cases Generation}
\label{sec:exp1}

This experiment evaluates the overall effectiveness and efficiency of \ourtool\ in the financial domain.

\noindent
\textbf{Competitors.}
We compare \ourtool\ against three representative categories of baselines: (1) \textit{Domain Experts}, consisting of three senior testing engineers with 3-5 years of financial experience; (2) \textit{LLM4Fin} \citep{xue2024llm4fin}, a SOTA hybrid regulatory test generation framework; and (3) \textit{End-to-End (E2E) LLMs}, represented by GPT-5, Grok-4, and DeepSeek-R1, which directly generate test cases from regulations using prompt engineering.

\noindent
\textbf{Datasets.} 
Due to the limited availability of public datasets with sufficient number and coverage, we constructed six financial datasets based on official regulatory documents from financial exchanges, paired with anonymized real-world test cases provided by industry partners.
The datasets span multiple regulatory frameworks and languages.
An overview is given in Table~\ref{tab:exp1_dataset}.
Notably, \textit{Datasets 4} and \textit{5} follow the evaluation setup of LLM4Fin, but use complete regulatory documents instead of 10--20 manually selected rules, resulting in a more realistic and challenging evaluation setting.

\begin{table}
    \setlength{\tabcolsep}{4pt}
    \caption{Details of the six financial evaluation datasets for Exp. I. \#X means the number of X, Req. means testable requirement, and TC means test case.}
    \label{tab:exp1_dataset}
    \centering
    \resizebox{\linewidth}{!}{
    \begin{tabular}{|c|p{5.3cm}|c|c|c|}
    \hline
        \textbf{Dataset} & \textbf{Document} & \textbf{\# Rule} & \textbf{\# Req.} & \textbf{\# TC} \\
        \hline
        \multirow{3}{*}{1} & New York Stock Exchange Auction Market Bids and Offers Rules \citep{dataset1} & \multirow{3}{*}{142} & \multirow{3}{*}{114} & \multirow{3}{*}{927} \\
        \hline
        \multirow{2}{*}{2} & The Nasdaq Stock Market Options Trading Rules \citep{dataset2} & \multirow{2}{*}{289} & \multirow{2}{*}{276} & \multirow{2}{*}{1465} \\
        \hline
        \multirow{2}{*}{3} & Hong Kong Stock Exchange Trading Rules \citep{dataset3} & \multirow{2}{*}{129} & \multirow{2}{*}{292} & \multirow{2}{*}{2128} \\
        \hline
        \multirow{2}{*}{4} & Shanghai Stock Exchange Trading Rules \citep{dataset4} & \multirow{2}{*}{212} & \multirow{2}{*}{195} & \multirow{2}{*}{682} \\
        \hline
        \multirow{2}{*}{5} & Shenzhen Stock Exchange Bond Trading Rules \citep{dataset5} & \multirow{2}{*}{190} & \multirow{2}{*}{167} & \multirow{2}{*}{1389} \\
        \hline
        \multirow{2}{*}{6} & Tokyo Stock Exchange Buying and Selling Rules \citep{dataset6} & \multirow{2}{*}{104} & \multirow{2}{*}{198} & \multirow{2}{*}{1020} \\
        \hline
    \end{tabular}
    }
\end{table}

\begin{table*}
	\setlength{\tabcolsep}{8.5pt}
    \centering
        \caption{Comparison of Precision (\%), Recall (\%), F1 (\%), and Business Scenario Coverage (BSC, \%) of test cases generated by Experts, LLM4Fin, End-to-end LLMs, and \ourtool\ on the six evaluation datasets.}
        \vspace{-2mm}
    \label{tab:exp1}
    \resizebox{\linewidth}{!}{
    \begin{tabular}{|c|llll|llll|llll|llll|}
    	\hline
    	\multirow{2}{*}{\textbf{Datasets}} & \multicolumn{4}{c|}{\textbf{Experts}} & \multicolumn{4}{c|}{\textbf{LLM4Fin}} & \multicolumn{4}{c|}{\textbf{End-to-end LLMs}} &  \multicolumn{4}{c|}{\textbf{\ourtool}}  \\ \cline{2-17}
    	              & Pre. & Rec. & F1 & BSC & Pre. & Rec. & F1 & BSC & Pre. & Rec. & F1 & BSC & Pre. & Rec. & F1 & BSC \\ \hline
    	            \textbf{1} & 83.7 & \textbf{96.3} & 89.6 & 90.0 & 80.3 & 70.2 & 74.9 & 78.1 & 7.1 & 41.0 & 11.8 & 32.2 & \textbf{95.0} & 95.8 & \textbf{95.4} & \textbf{90.2} \\ 
    	            \textbf{2} & \textbf{90.8} & \textbf{91.5} & \textbf{91.1} & 85.3 & 52.2 & 72.8 & 60.8 & 76.6 & 8.2 & 21.1 & 11.7 & 29.8 & 85.0 & 84.4 & 84.7 & \textbf{88.4} \\ 
                    \textbf{3} & 95.1 & \textbf{98.9} & \textbf{96.9} & \textbf{92.3} & 74.3 & 77.0 & 75.6 & 79.4 & 11.9 & 33.6 & 16.8 & 42.2 & \textbf{96.9} & 93.4 & 95.0 & 91.1 \\ 
                    \textbf{4} & \textbf{93.5} & 93.7 & \textbf{93.6} & 81.1 & 59.9 & 82.1 & 69.3 & 72.8 & 27.0 & 46.7 & 33.6 & 40.0 & 79.1 & \textbf{98.7} & 87.8 & \textbf{83.1} \\ 
                    \textbf{5} & \textbf{97.3} & 96.0 & \textbf{96.6} & 89.1 & 61.2 & 87.7 & 72.1 & 83.7 & 25.4 & 49.8 & 33.2 & 38.0 & 95.4 & \textbf{97.4} & 96.4 & \textbf{89.5} \\ 
                    \textbf{6} & \textbf{94.3} & \textbf{97.1} & \textbf{95.7} & \textbf{86.1} & 83.7 & 41.8 & 55.7 & 48.6 & 19.1 & 36.5 & 24.6 & 31.0 & 92.1 & 89.7 & 90.9 & 76.9 \\ \hline
                    \textbf{Average} & 92.4 & 95.6 & 93.9 & 87.3 & 68.6 & 71.9 & 68.1 & 73.2 & 16.4 & 38.1 & 21.9 & 35.5 & 90.6 & 93.2 & 91.7 & 86.5 \\
                    \textbf{Variance} & 18.9 & 5.6 & 7.7 & 13.2 & 132.8 & 215.2 & 54.6 & 131.7 & 62.2 & 88.3 & 83.8 & 22.4 & 41.4 & 24.0 & 18.6 & 25.0 \\ \hline
    \end{tabular}
    }
\end{table*}

\noindent
\textbf{Metrics.}
Due to strict security constraints in financial institutions, executable system code is unavailable. Following prior work~\citep{tufano2020unit,fatima2022flakify}, we use anonymized real-world test cases adopted in practice as ground truth.
Effectiveness is evaluated using \textit{Precision}, \textit{Recall}, and \textit{F1} with the adopted test cases, as well as \textit{Business Scenario Coverage (BSC)}~\citep{xue2024llm4fin} to assess regulatory requirement coverage.
Efficiency is measured by \textit{Time Consumption} and \textit{Token Usage}.
We report the time required for regulatory analysis (if applicable), test generation, and expert review to achieve 100\% correctness.
For LLM-based methods, we additionally report prompt and completion token usage and the corresponding monetary cost.

\noindent
\textbf{Experimental Procedure.}
Each of the six datasets was provided to the three domain experts, who manually authored compliance test cases.
For E2E LLMs, prompts were designed following SOTA prompt engineering practices (see Appendix~\ref{sec:app_prompt_llm}).
LLM4Fin was executed fully automatically using its default configuration.
For \ourtool, test cases were generated following the workflow described in Section \ref{sec:method}.
Reported results for domain experts are averaged across the three individuals, while results for E2E LLMs and \ourtool\ are averaged over multiple runs using GPT-5, Grok-4, and DeepSeek-R1.


\noindent
\textbf{Experimental Results.}
The effectiveness results are summarized in Table~\ref{tab:exp1}.
Domain experts achieve the highest performance, while \ourtool\ attains comparable results and substantially outperforms all automated baselines.
Across the six datasets, \ourtool\ balances precision and recall, achieving an average F1 of 91.7\% and BSC of 86.5\%, improving over LLM4Fin by 34.7\% and 18.2\%, respectively.
LLM4Fin underperforms due to its reliance on smaller BERT-based models with limited domain knowledge, which hinders its ability to capture complex and implicit regulatory logic.
In contrast, E2E LLMs suffer from hallucinations, reasoning errors, and output instability, resulting in lower accuracy and coverage.
Although \ourtool\ exhibits slightly higher variance than human experts due to its reliance on probabilistic LLM outputs, its variance is markedly lower than that of E2E LLMs and LLM4Fin, demonstrating that the proposed multi-LLM \strategy\ strategy effectively mitigates hallucination and randomness.

\begin{figure}
    \centering
    \includegraphics[width=\linewidth]{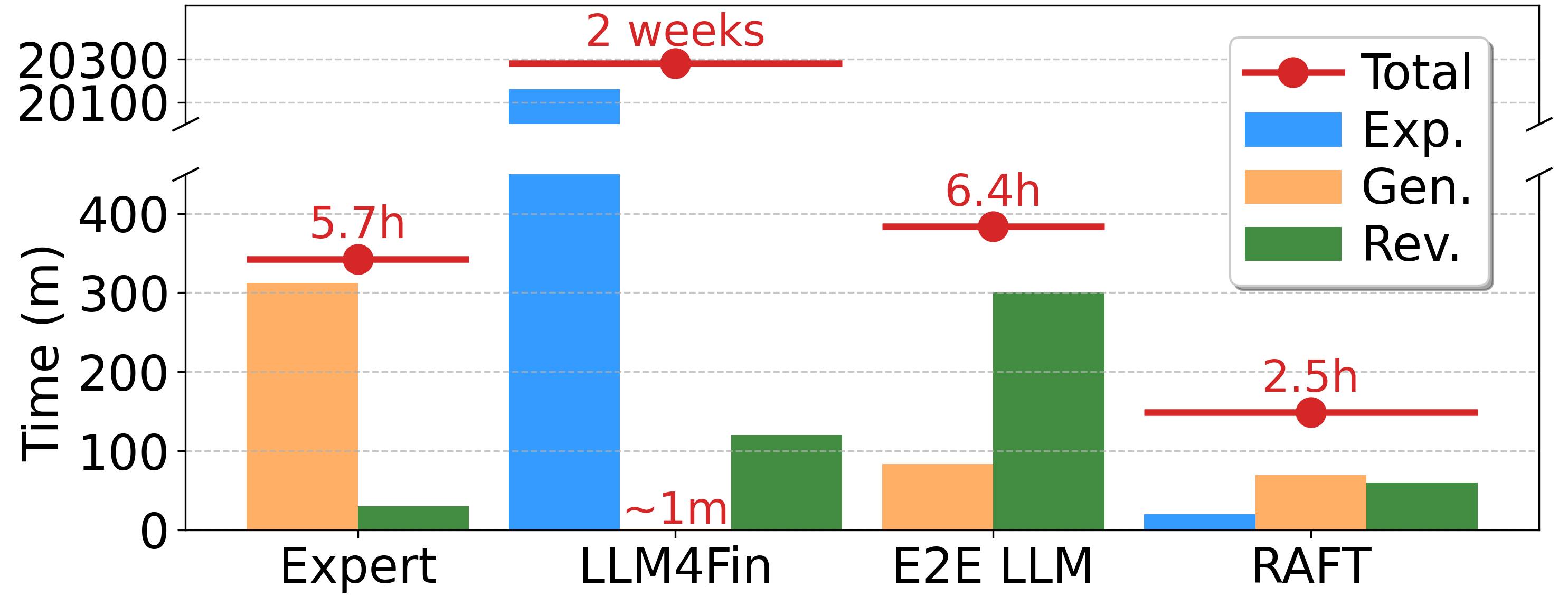}
    \caption{Comparison of the time consumption on knowledge explication (Exp.), test generation (Gen.), and expert review (Rev.) on processing the six datasets.}
    \label{fig:exp1_time}
\end{figure}

Efficiency results, shown in Figure~\ref{fig:exp1_time} and Table~\ref{tab:exp1_token}, further highlight the practical advantages of \ourtool.
Test generation by domain experts and LLM4Fin requires substantial manual effort, ranging from 5.7 hours to 2 weeks.
In contrast, \ourtool\ significantly reduces end-to-end time, achieving a 2.3$\times$ speedup over domain experts and more than a 134$\times$ speedup over LLM4Fin.
Even compared to E2E LLMs, \ourtool\ provides a 156\% speedup, primarily because its high-accuracy test cases reduce expert review time from 2--4 hours to approximately 1 hour.
From a cost perspective, \ourtool\ is also more economical, as it emphasizes prompt tokens over expensive completion tokens.
Although LLM4Fin minimizes token usage through local deployment, this advantage comes at the expense of reduced test quality due to limited model capacity.



\noindent
\textbf{Conclusion:}
\ourtool\ achieves expert-level effectiveness, significantly improving efficiency and reducing cost, demonstrating its practicality for real-world regulatory compliance testing.

\begin{table}
    \setlength{\tabcolsep}{7pt}
    \caption{Comparison of the average token usage and monetary cost for each LLM-based approach.}
    \label{tab:exp1_token}
    \centering
    \resizebox{\linewidth}{!}{
    \begin{tabular}{|c|c|c|c|}
    \hline
    \textbf{Method} & \textbf{Prompt Tokens} & \textbf{Completion Tokens} & \textbf{Cost (\$)} \\
    \hline
    LLM4Fin & 227.3K & 114.8K & 0 \\
    E2E LLM & 115.8K & 1328.6K & 12.34 \\
    \ourtool\ & 1047.0K & 128.3K & 2.84 \\
    \hline
    \end{tabular}
    }
\end{table}

\begin{figure*}
    \centering
    \includegraphics[width=\linewidth]{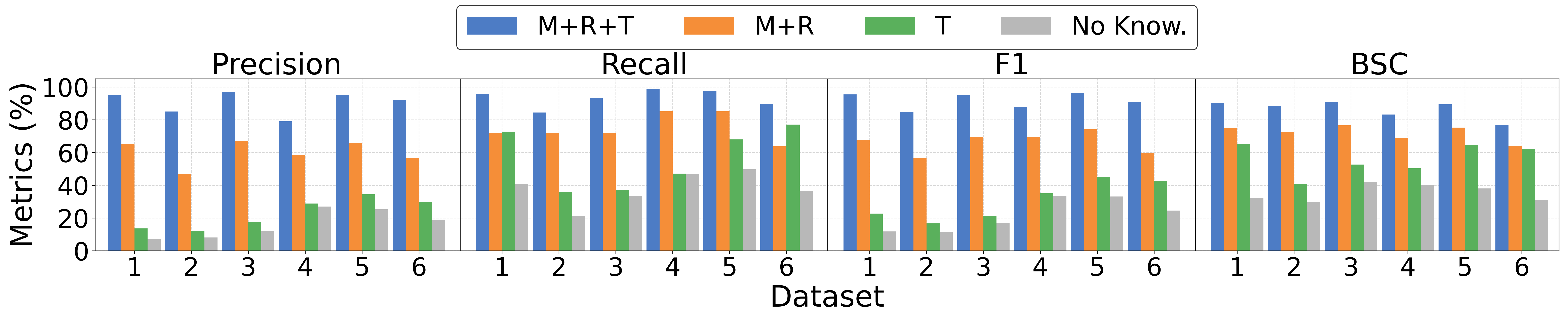}
    \caption{Impact of regulatory knowledge components ($\mathcal{M}, \mathcal{R}, \mathcal{T}$) on the quality of generated test cases.}
    \label{fig:exp2}
\end{figure*}

\begin{figure*}
    \centering
    \includegraphics[width=\linewidth]{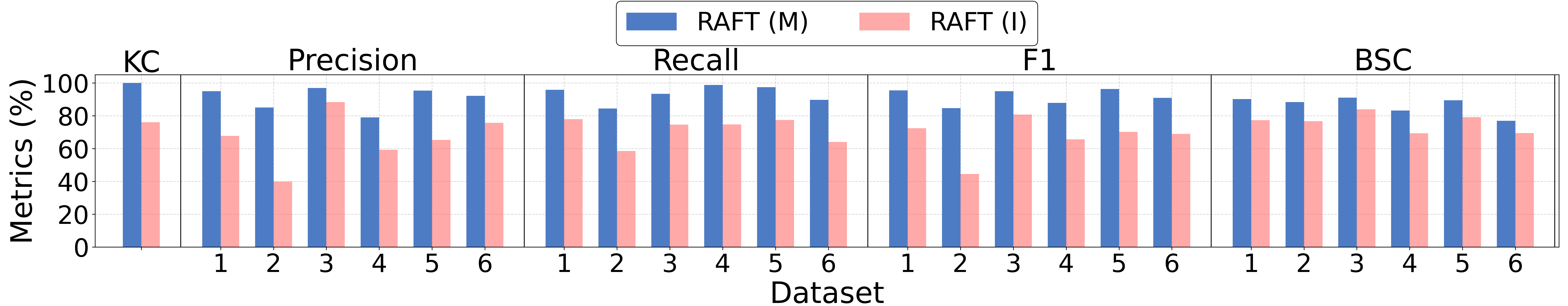}
    \caption{Comparison between our \strategy\ strategy (\ourtool\ (M)) and individual LLMs (\ourtool\ (I)) on knowledge coverage (KC) and metrics of generated test cases.}
    \label{fig:exp2_2}
\end{figure*}

\subsection{Experiment II: Ablation Study}
\label{sec:exp2}

\noindent
\textbf{Experimental Design.}
We evaluate the necessity of explicated regulatory knowledge by comparing the full \ourtool, including meta-model $\mathcal{M}$, requirements representation $\mathcal{R}$, and testability constraints $\mathcal{T}$, against three degraded variants: 
(1) w/o $\mathcal{T}$, which generates test cases without filtering non-testable requirements; (2) w/o $\mathcal{M}$ \& $\mathcal{R}$, which generates test cases directly from rules after testability assessment; and (3) w/o all: which relies on E2E LLMs without explicit regulatory knowledge.

To assess the proposed \strategy\ strategy, we further compare the full \ourtool\ using GPT-5, Grok-4, and DeepSeek-R1 with variants that perform knowledge explication using them separately.
All datasets, metrics, and procedures follow Experiment I.

\noindent
\textbf{Experimental Result.}
As shown in Figure~\ref{fig:exp2}, removing $\mathcal{T}$ causes a notable precision drop of 33.6\% due to noise introduced by non-testable requirements.
Removing both $\mathcal{M}$ and $\mathcal{R}$ leads to a more severe degradation, with F1 decreasing by 66.5\%, indicating that LLMs struggle to capture regulatory semantics and implicit logic without explicit guidance.
The poorest performance of the no-knowledge variant confirms that integrating $\mathcal{M}$, $\mathcal{R}$, and $\mathcal{T}$ is essential for high-quality test generation.

Figure~\ref{fig:exp2_2} shows that the \strategy\ strategy consistently outperforms the single-LLM variants.
While individual LLMs are prone to hallucination, resulting in incomplete knowledge explication and degraded test quality, the consensus-based strategy aggregates complementary insights to produce more precise and robust regulatory knowledge, thereby improving the quality of generated test cases.

\noindent
\textbf{Conclusion:}
Explicit regulatory knowledge ($\mathcal{M}$, $\mathcal{R}$, $\mathcal{T}$) is essential for effective compliance test generation, and the proposed \strategy\ strategy further enhances it through a more comprehensive knowledge explication.

\begin{table}
    \setlength{\tabcolsep}{3pt}
    \caption{Details of the five evaluation datasets on automotive and power domains for Exp. III.}
    \label{tab:exp2_dataset}
    \centering
    \resizebox{\linewidth}{!}{
    \begin{tabular}{|c|c|p{5cm}|c|c|c|}
    \hline
    \textbf{Dataset} & \textbf{Domain} & \textbf{Document} & \textbf{\# Rule} & \textbf{\# Req.} & \textbf{\# TC} \\
        \hline
        \multirow{2}{*}{7} & \multirow{6}{*}{Automotive} & 2025 New Jersey Driver Manual \citep{dataset7} & \multirow{2}{*}{47} & \multirow{2}{*}{427} & \multirow{2}{*}{1570} \\
        \cline{1-1}\cline{3-6}
        \multirow{2}{*}{8} & & Road Traffic Safety Law of PRC \citep{dataset8} & \multirow{2}{*}{171} & \multirow{2}{*}{613} & \multirow{2}{*}{2124} \\
        \cline{1-1}\cline{3-6}
        \multirow{2}{*}{9} & & Cambodian Road Traffic Law \citep{dataset9} & \multirow{2}{*}{93} & \multirow{2}{*}{808} & \multirow{2}{*}{3132} \\
        \hline
        \multirow{2}{*}{10} & \multirow{3}{*}{Power} & EN 50549-1:2019 \citep{dataset10} & \multirow{2}{*}{75} & \multirow{2}{*}{362} & \multirow{2}{*}{703} \\
        \cline{1-1}\cline{3-6}
        \multirow{1}{*}{11} & & GB/T 19964-2024 \citep{dataset11} & \multirow{1}{*}{189} & \multirow{1}{*}{195} & \multirow{1}{*}{1311} \\
        \hline
    \end{tabular}
    }
\end{table}

\subsection{Experiment III: Method Generalizability}
\label{sec:exp3}

\noindent
\textbf{Experimental Design.}
We evaluate the generalizability of \ourtool\ in the automotive and power domains.
Five datasets were constructed, covering three traffic regulations and two power grid standards, each paired with real-world industrial test cases, as summarized in Table~\ref{tab:exp2_dataset}.
Together with the financial datasets from Experiment~I, this evaluation spans three regulated domains.

\begin{figure}
    \centering
    \includegraphics[width=\linewidth]{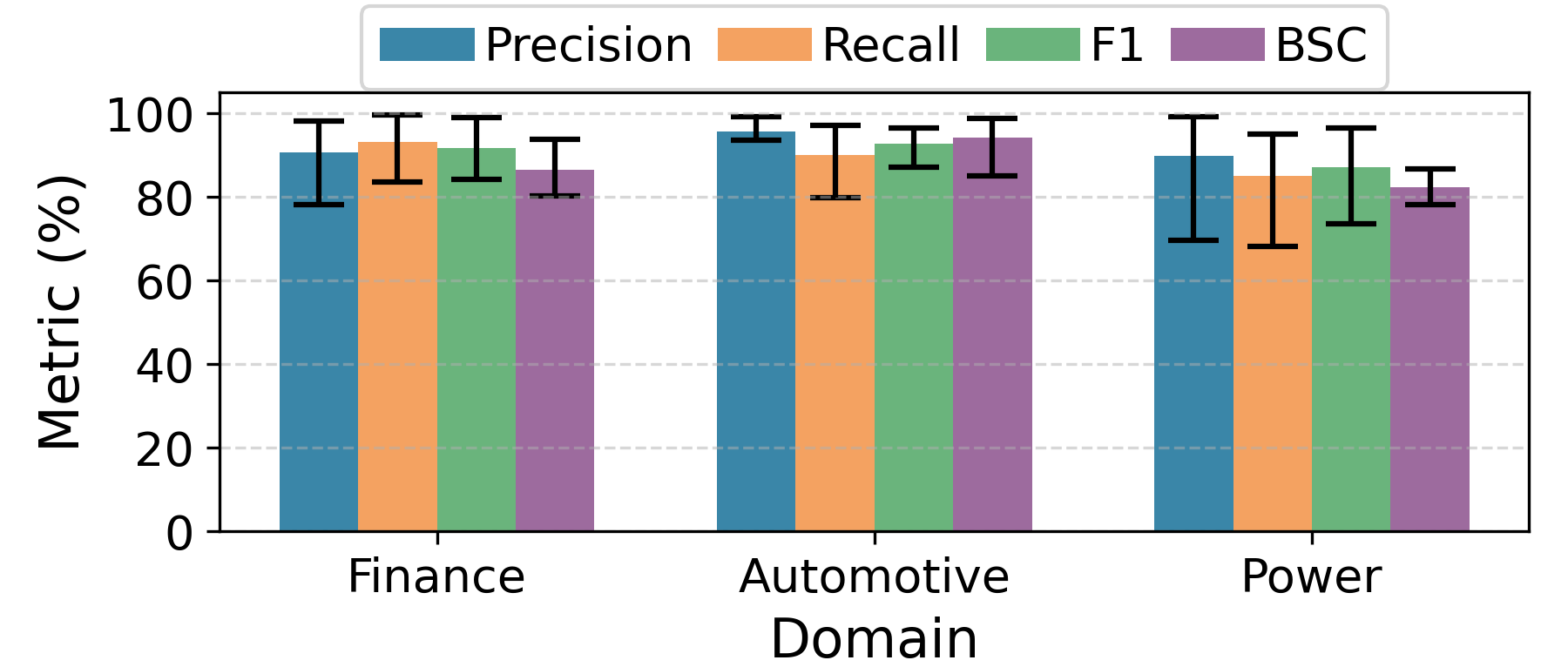}
    \caption{Average performance of test cases generated by our approach across three industrial domains.}
    \label{fig:exp3}
\end{figure}

\noindent
\textbf{Experimental Results.}
The results are shown in Figure \ref{fig:exp3}. \ourtool\ demonstrates consistently strong performance across all 11 datasets.
In the automotive and power domains, it achieves average F1/BSC scores of 92.7\%/87.2\% and 94.3\%/82.3\%, respectively, comparable to those in the financial domain.
Its performance remains stable across domains, with variance below 20 in most cases.
These results indicate that the proposed regulatory knowledge explication and test generation approach generalizes well across diverse regulatory contexts.



\noindent
\textbf{Conclusion:}
Beyond the financial domain, \ourtool\ maintains high performance in both automotive and power domains, demonstrating its strong cross-domain generalizability.

\section{Conclusion}
\label{sec:conclusion}


In this paper, we propose \ourtool, a framework that automatically explicates tacit regulatory knowledge from LLMs into structured artifacts to drive high-quality test case generation. Experimental results across financial, automotive, and power domains demonstrate that \ourtool\ achieves expert-level effectiveness, high efficiency, and strong cross-domain generalizability. By shifting from manual modeling to knowledge-driven automation, our approach provides a scalable and reliable solution for compliance testing in highly regulated domains.

\section*{Limitations}

\noindent
\textbf{Dependence on the Quality and Coverage of External Knowledge Sources.}
Although the proposed framework supplies knowledge for LLMs by incorporating regulatory documents and historical test cases through RAG, its performance still depends on the quality, completeness, and representativeness of these external knowledge sources. In domains where regulations are ambiguous, rapidly evolving, or insufficiently documented, the extracted domain models and requirements representations may be incomplete or biased, thereby affecting downstream test generation. Addressing this limitation requires continuous maintenance and supplementation of the knowledge base, and mechanisms for detecting and resolving gaps or inconsistencies in regulatory knowledge.

\noindent
\textbf{Limited Formal Guarantees on Model Soundness and Completeness.}
While the collective wisdom strategy and model-based generation improve controllability compared to end-to-end LLM approaches, the automatically constructed meta-models and requirements representations lack formal guarantees of soundness or completeness with respect to the original regulations. Subtle semantic nuances, implicit assumptions, or exceptional regulatory clauses may still be partially or inaccurately captured. As a result, human expert review remains necessary, especially for safety-critical or legally sensitive scenarios. Future work could explore integrating formal verification techniques or regulation-specific logical formalisms to further strengthen the theoretical guarantees of the generated models and test cases.


\bibliography{custom}

\appendix
\section{Appendix}
\label{sec:appendix}

\subsection{Prompt Design in Our Framework}
\label{sec:app_prompt_ours}

\subsubsection{Meta-model Construction}

The prompt design theories and techniques used for constructing financial meta-models from regulatory texts and test cases have been detailed in Section \ref{sec:know_ext}. It integrates role-based instructions, structural constraints, and staged outputs to support the abstraction of domain knowledge into a formal and test-oriented representation. External financial regulations and test cases are provided as supplementary knowledge to enhance semantic completeness. 

The specific prompts applied are shown below. The proposed prompt design has been empirically validated across multiple domains, including finance, automotive, and energy, and evaluated on diverse LLMs such as GPT, Grok, and DeepSeek, demonstrating its domain-agnostic nature and independence from specific model implementations.

\begin{markdowncode} 
You are a software requirements modeling and testing expert. I will provide two types of inputs:
1. A domain rule document;
2. Corresponding partial test cases.

Your task is to construct a **domain meta-model** for the rules of the domain. The meta-model should describe the syntax of rules in a class diagram-style tree structure. The construction of the meta-model must be completed in two steps: first, extract and confirm rule elements (15-20 leaf elements); second, based on these elements, build a three-layer meta-model and establish relationships between these elements and Precondition, Operation, and ExpectedResult.
Please strictly follow the following specifications and output format: strict, complete, and machine-readable:

### Meta-Model Construction Specifications (Must be followed)
1. **Overall Structure**
   * The meta-model should be a tree structure in class diagram style (classes and relationships only, without attributes or methods).
   * Must adopt a three-layer structure: top layer (1 node) -> middle layer (fixed 3 nodes) -> bottom layer (n leaf nodes).
   * The top-layer node is fixed as `Rule` (single class).
   * The middle layer is fixed as three classes: `Precondition`, `Operation`, `ExpectedResult`.
   * The bottom layer consists of minimal rule elements (leaf nodes).

2. **Leaf Node (Rule Element) Requirements**
   * Must be testable, quantifiable, and concrete rule elements.
   * The total number of elements must be around 15. High-frequency/core elements are listed individually; semantically overlapping elements are merged; scattered low-frequency elements are merged into a general leaf `Others`. Note: Precondition, Operation, and ExpectedResult should all include `Others`.
   * Leaf nodes should be attributes (keys), and must not include specific values or examples (e.g., "interest rate" is correct, "annual interest rate 5%" is incorrect).
   * Preferably extracted from test cases and supplemented by rule documents. Elements presented in test cases as ``RuleElement1/RuleElement2'' should be merged as a single element.
   * Elements should be reusable within the domain and cover common elements across different domain rule documents.

3. **Relationships**
   * Relationship types and applicable scenarios:
     * `contains` - composition relationship of middle-layer nodes to bottom-layer elements
     * `constrains` - constraint relationship of elements to nodes/other elements
     * `dependsOn` - prerequisite relationship of elements at runtime
     * `triggers` - operation triggering expected result
     * Other relationships
   * Every relationship must be labeled with its name in PlantUML.
   * Bottom-layer elements can only establish relationships with one of the three middle-layer nodes, and the relationship type must match the element's semantics. Carefully consider whether each rule element belongs to Precondition, Operation, or ExpectedResult.

4. **Two-Step Output Requirements**
   * **Step 1 - Rule Element List**: Output a list of approximately 15 leaf classes (class names in UpperCamelCase), each with a semantic description within 10 words.
   * **Step 2 - Meta-Model**: Construct a complete three-layer class diagram in PlantUML:
     * Top layer `Rule`;
     * Middle layer `Precondition`, `Operation`, `ExpectedResult`;
     * Bottom layer directly referencing the leaf classes from Step 1, without renaming.
   * In Step 2, clearly indicate the direction and type of all relationships, ensuring a tree hierarchy and logical consistency.

5. **Naming and Style**
   * All class names use UpperCamelCase (e.g., `AccountStatus` rather than `account_status`).
   * Class bodies remain empty (class name only).
   * Relationship names use lowercase verbs (e.g., `contains`, `triggers`).

### Output (Mandatory)
* The rule element list should be output as a numbered list; the meta-model should be output as PlantUML code.
* The number of leaf nodes must be strictly around 15; the leaf class names in Step 1 and Step 2 must match exactly (including capitalization).
* If you are unsure of which elements to include, refer to the keys in the test cases. Elements must be concise, representative, and high-frequency. For example, only one Actor should exist as a primary entity; having X Actor, Y Actor, etc., would be too repetitive.
\end{markdowncode}

\subsubsection{Requirements Representation Formalization}

Requirement Representation Formalization defines a formal, machine-readable requirement representation that captures the essential semantics of domain rules while enabling consistent parsing and downstream analysis.
Given natural-language rules, example test cases, and a domain meta-model, this step formalizes informal requirements into a formal representation that bridges free-text specifications and executable artifacts, without yet enforcing testability constraints.

The formalized requirement representation consists of two core components: Symbols and Syntax.
Symbols define the vocabulary and semantic roles of elements used in requirement expressions, while Syntax specifies the hierarchical structure and composition rules for combining these elements into well-formed requirements.
This representation is domain-agnostic and independent of any specific modeling language or testing framework, enabling reuse across different regulatory domains and LLM backends.

\begin{markdowncode}
You are a software requirements modeling expert.

Your task is to define a formal requirement representation language for describing domain rules in a structured and machine-readable form.

This representation aims to formalize natural language rules or requirement documents so that they can be consistently parsed, analyzed, and used as a basis for further validation and testing.

### Input
You will receive the following three types of input:
1. **Rule/Requirement Document** - natural language descriptions of domain rules or system behavior;
2. **Test Cases** - example test cases that illustrate the intended behavior of some rules;
3. **Meta-Model Syntax Structure** - defining the core elements of rules and their hierarchical relationships.

### Task Objective
Based on the above inputs, define a complete formal Requirement Representation.
The language definition includes two parts: **Symbols** and **Syntax**.

### **Symbol System**
* Based on the given meta-model, define the types of symbols used in the language and their scope, including:
  * Logical Symbols
  * Comparison Symbols
  * Domain Symbols
* Clearly specify the syntactic position and semantic role of each type of symbol.

### **Syntax**
* Based on the given meta-model, define the core components of the language and their hierarchical composition rules;
* The value of each element should be a string or number, not an enum;
* Use formal BNF to define the language structure;
* The syntax should be simple and explicit, while supporting complex rule expressions (e.g., AND/OR/NOT, nested conditions, composite actions);
* Condition, Action, and ExpectedOutcome should follow consistent structural patterns.

### **Output Requirements**
Please output:
1. The formal definition of Symbols and Syntax;
2. A complete definition of the requirement representation;
3. **An example** showing how a natural language rule is transformed into an instance of this representation.

### Meta-Model (PlantUML Representation)
{}

\end{markdowncode}

\subsubsection{Testability Constraints Specification}

Testability Constraint Specification defines a set of explicit constraints over the formalized requirement representation to determine whether a requirement is testable.
This step focuses on identifying structural, semantic, and data-related conditions that must be satisfied for a requirement to support automated test generation and verification.

Testability constraints are specified independently of requirement syntax and are applied as a validation layer over the formal requirement representation.
These constraints capture necessary conditions such as observability, determinism, input completeness, and verifiable outcomes, and are formalized using OCL to enable automated checking.

\begin{markdowncode}

You are a software testing and requirements validation expert.

Your task is to define a set of formal testability constraints for a given formal requirement representation.

These constraints are intended to determine whether a requirement instance is testable, i.e., whether it can be unambiguously verified and used for automated test generation.

### Input
You will receive:
1. **Formal Requirement Representation Definition** - including Symbols and Syntax.

### Task Objective
Based on the above inputs, define a comprehensive set of **Testability Constraints**.

### **Testability Constraints**
* Identify the necessary and sufficient conditions for a requirement to be testable, considering:
  * Completeness of conditions and actions;
  * Observability and verifiability of expected outcomes;
  * Absence of ambiguity or underspecification;
  * Compatibility with available test inputs and outputs.
* Formalize these constraints using **OCL** over the requirement representation model.

### **Output Requirements**
Please output:
1. A complete set of testability constraints expressed in OCL;
2. A clear explanation of each constraint and its rationale;
3. **An example** demonstrating how a formalized requirement is checked against these constraints and classified as testable or non-testable.


### Formal Requirements Representation
{}

\end{markdowncode}

\subsubsection{Formal Requirement Generation}

This section presents formal requirement generation, which transforms natural language rules into formal, testable requirements based on the testable requirements representation. The task is guided by prompts specifying the expert role, task objective, inputs, output format, and language constraints. While earlier methods required model training \citep{xue2024llm4fin}, modern large language models can achieve strong results using prompt engineering with few-shot examples alone. An example prompt in the financial domain is provided below.

\begin{markdowncode}
You are a software requirements modeling and testing expert, proficient in requirements engineering, formal modeling, and test case generation. You are familiar with testable requirements representation, which can accurately represent rules described in natural language as formal requirements that are executable, verifiable, and testable.

### Task Description
I will provide one or more rules described in natural language, and your task is to convert them into the Testable Requirement Language.

### Input
1. Definition of the Testable Requirement Language (provided below).
2. Natural language rules to be converted, provided one at a time in multiple interactions.

### Output
Formal Requirements

### Generation Requirements
1. The output testable requirements must conform to the elements and syntax of the representation, and must not exceed the defined scope.
2. If a rule can be interpreted in multiple ways, provide the best expression.
3. TRL must accurately express the meaning of the rule and fully include all elements of the rule; nothing should be omitted.
4. Only output the testable requirement; do not output extra content, and do not evaluate testability at this step (this will be handled by other tools later).

### Testable Requirement Language Definition
#### Symbol Definition
{}

#### Syntax Definition
{}

#### Other Requirements
1. Generally, strictly follow the defined symbols and syntax when writing testable requirements.
2. Except for cases requiring modulo calculations, all expressions must follow the key-op-value triple format; binary, quadruple, or quintuple formats are prohibited.
3. Must accurately express the rule meaning and fully include all elements (subject, verb, object, events, and domain-specific elements); redundancy is allowed, omission is not.
4. Do not overuse parentheses; use only when necessary (e.g., with or, not).
5. Remain faithful to the original text; only words present in the original text may appear. No invented words are allowed.

### Examples
* Example 1
  * Input: Unless otherwise specified by the exchange, the bond trading submission quantity shall comply with the following requirements: (1) for matching execution, the bond cash submission quantity shall be 100,000 yuan or multiples thereof; for sales of less than 100,000 yuan, submit in a single order; for general pledged repo of bonds, the submission quantity shall be 1,000 yuan or multiples thereof; (2) for click execution, the submission quantity shall be 100,000 yuan or multiples thereof; (3) for inquiry or competitive purchase execution, the submission quantity shall be no less than 100,000 yuan and in multiples of 1,000 yuan; (4) for negotiated execution, the bond cash submission quantity shall be no less than 1,000 yuan and in multiples of 100 yuan; general pledged repo submission quantity shall be 1,000 yuan or multiples thereof.
  * Output:
```
rule 1
if TradingInstrument = Bond and TradingMethod = MatchingExecution and BondType = CashBond and Quantity % 100000 == 0
then Result = Success
rule 2
if TradingInstrument = Bond and TradingMethod = MatchingExecution and BondType = CashBond and Action = Sell and Balance < 100000 and Action = SingleSubmission
then Result = Success
rule 3
if TradingInstrument = Bond and TradingMethod = MatchingExecution and BondType = GeneralPledgedRepo and Quantity % 1000 == 0
then Result = Success
rule 4
if TradingInstrument = Bond and TradingMethod = ClickExecution and Quantity % 100000 == 0
then Result = Success
rule 5
if TradingInstrument = Bond and (TradingMethod = InquiryExecution or TradingMethod = CompetitivePurchase) and Quantity >= 100000 and Quantity % 1000 == 0
then Result = Success
rule 6
if TradingInstrument = Bond and TradingMethod = NegotiatedExecution and BondType = CashBond and Quantity >= 1000 and Quantity % 100 == 0
then Result = Success
rule 7
if TradingInstrument = Bond and TradingMethod = NegotiatedExecution and BondType = GeneralPledgedRepo and Quantity % 1000 == 0
then Result = Success
```

* Example 2
  * Input: If both parties agree to a manual method, and the client wishes to continue participating in the next period on the repo maturity date, they shall re-issue the initial order to the securities company.
  * Output:
```
rule 1
if Actor = BothParties and Action = Agree and Constraint = ManualMethod and Actor = Client and Time = RepoMaturityDate and Action = WishToContinue and OperationPart = NextPeriodTrade and OperationTarget = SecuritiesCompany and Action = Issue and OperationPart = InitialOrder
then Result = Success
```
\end{markdowncode}

\subsubsection{Testability Determination}
This section covers testability determination, which evaluates whether a requirement satisfies specified testability constraints. Using few-shot prompts, the model receives a natural language rule, its formal requirement, and a single constraint per evaluation. The process checks constraints iteratively: if the constraint is satisfied, it proceeds to the next; if not, it stops. This leverages large language models’ reasoning without additional training.

\begin{markdowncode}
You are a software requirements modeling and testing expert, specialized in evaluating whether a natural language rule and its formalized requirement constitute a testable requirement.
Now, you need to determine whether a given rule is a **Testable Requirement**.
The rule inputs include:
1. Natural Language Rule
2. Formalized Rule
You need to assess whether the rule meets both requirement validity and testability criteria, and output:
* `True` (testable requirement) or `False` (non-testable or non-requirement)
* If `False`, provide a brief explanation.

### **Input Format**
For example, a rule in natural language:
The trading system shall complete matching within 5 seconds after receiving a valid order.
Its corresponding formalized expression:
```
rule 1
if Actor = TradingSystem and Event = ReceivedOrder and Time = within5s and Action = CompleteMatching
then Result = Success
```


### **Evaluation Criteria**
{}


### **Output Format**
Output `true` or `false` to indicate whether the rule is a testable requirement.
If the rule is not testable, provide a one-sentence explanation.

### **Reasoning Strategy (Thinking Sequence)**
1. First, determine if it is a requirement (exclude background, definitions, etc.).
2. Then, check each of the five testability constraints one by one.
3. If all are satisfied, output `True`; otherwise, output `False`.
4. The explanation should be concise and focused on verifiability.

### **Some Intuitive Guidelines**
1. Statements like ``may'' are considered acceptable, as doing or not doing does not affect testability determination.
2. All elements of a testable requirement must be deterministic; expressions such as ``other'', ``unless otherwise specified'', or ``submission time (instead of a specific 9:00-10:00)'' are not allowed.
3. Testable requirements cannot contain references (e.g., Article A, previous chapter, or earlier rules).
4. The formalized rule has already undergone some testability processing and is more testable than the original text. Evaluation mainly focuses on whether the formalized rule meets testability requirements; the natural language text serves as a reference.
Note: All testability determinations use strict OCL constraint patterns. No contextual assumptions are made; the judgment is based on pure textual-level testability (must be independent and directly verifiable).

### **Example**
#### Input:
When the system receives a user payment instruction, it should immediately deduct the corresponding account balance and return the transaction result status.
```
rule 1
if Actor = System and Event = ReceivedUserPaymentInstruction and Action = Deduct and OperationPart = AccountBalance and Action = Return and OperationPart = TransactionResultStatus
then Result = Success and ResultStatus = TransactionSuccess
```
#### Output:
true
\end{markdowncode}

\subsection{Extracted Regulatory Knowledge for the Financial Domain}
\label{sec:app_knowledge}

This chapter presents the domain meta-model and the testable requirements representation for the financial domain.

\subsubsection{Meta-Model}

\begin{artifact}
@startuml
skinparam defaultFontSize 20
skinparam ranksep 20
skinparam nodesep 20

class TradingRule

class Condition
class Operator
class Indicator
class Parameter
class Action
class Signal
class RuleSet
class PositionSizingRule
class RiskManagementRule
class EntryRule
class ExitRule

TradingRule *-- Condition : conditions
TradingRule *-- Action : actions
TradingRule -- RuleSet : part of
RuleSet *-- TradingRule : contains

Condition *-- Operator
Condition *-- Indicator
Condition *-- Parameter

Action -- Signal

EntryRule --|> TradingRule
ExitRule --|> TradingRule
PositionSizingRule --|> TradingRule
RiskManagementRule --|> TradingRule
@enduml
\end{artifact}

\subsubsection{Formal Requirements Representation}
\begin{artifact}
### Symbols:
* Logical Symbols:
  * and, or, not
  * implicate
* Comparison Symbols
  * =, !=, >, >=, <, <=, in
* Domain Symbols
  * Precondition: Actor, TradingInstrument, TradingMarket, Time, Constraint, Event
  * Action: Action, TradingDirection, TradingMethod, Quantity, Price, OperationPart, Status
  * ExpectedResult: ResultStatus, Result

### Syntax:
```BNF
Rule ::= "IF" <Precondition> "AND" <Operation> "THEN" <ExpectedOutcome>

Precondition ::= <AtomicPrecondition> | <CompoundPrecondition>
AtomicPrecondition ::= <PreconditionElement> <Comparator> <Value>
CompoundPrecondition ::= "(" <Precondition> ")" 
                    | <Precondition> "AND" <Precondition>
                    | <Precondition> "OR" <Precondition>
                    | "NOT" <Precondition>

Operation ::= <AtomicOperation> | <CompoundOperation>
AtomicOperation ::= <OperationElement> <Comparator> <Value>
CompoundOperation ::= "(" <Operation> ")" 
                 | <Operation> "AND" <Operation>
                 | <Operation> "OR" <Operation>
                 | "NOT" <Operation>

ExpectedOutcome ::= <AtomicOutcome> | <CompoundOutcome>
AtomicOutcome ::= <ResultElement> <Comparator> <Value>
CompoundOutcome ::= "(" <ExpectedOutcome> ")" 
                  | <ExpectedOutcome> "AND" <ExpectedOutcome>

PreconditionElement ::= "Actor" | "TradingInstrument" | "TradingMarket" 
                | "Time" | "Event" | "Constraint"
                
OperationElement ::= "Action" | "TradingDirection" | "TradingMethod" 
                   | "Quantity" | "Price" | "OperationPart" | "Status" | "Constraint"
                   
ResultElement ::= "ResultStatus" | "Result" | "Constraint"

Comparator ::= "=" | "!=" | ">" | "<" | ">=" | "<=" | "in"

Value ::= <StringLiteral> | <NumberLiteral> | <BooleanLiteral> | <TimeLiteral> | <TimeRangeSet>
StringLiteral ::= "\"" [^"]* "\""
NumberLiteral ::= [0-9]+ ("." [0-9]+)?
BooleanLiteral ::= "true" | "false"

TimeLiteral ::= "\"" [0-9]{2} ":" [0-9]{2} (":" [0-9]{2})? "\""  // Support minutes and seconds
TimeRange ::= <TimeLiteral> "-" <TimeLiteral>
TimeRangeSet ::= "[" <TimeRange> ("," <TimeRange>)* "]"  // Example: [10:00-12:00,13:00-14:00]
```
\end{artifact}

\subsubsection{Testability Constraints}
\begin{artifact}
1. Structural Completeness: Condition, Action, and Result must be non-empty
context Rule
```OCL
inv StructuralCompleteness:
    not self.Precondition.oclIsUndefined() and
    not self.Operation.oclIsUndefined() and
    not self.ExpectedResult.oclIsUndefined()
```
2. Deterministic Rule Elements: Precondition, Operation, and ExpectedResult elements must be concrete and deterministic
```OCL
context Rule
inv RuleElementDeterministic:
    self.Precondition->forAll(e | e.concrete() and e.deterministic())
    self.Operation->forAll(e | e.concrete() and e.deterministic())
    self.ExpectedResult->forAll(e | e.concrete() and e.deterministic())
```
3. Action Executability: Actions must be executable
```OCL
context Rule
inv ActionExecutable:
    self.Operation.Action.notEmpty() and
    self.Operation.Action.executable()
```
4. Expected Result Observability: Results must be observable
```OCL
context Rule
inv ExpectedResultObservable:
    self.ExpectedResult.Result.notEmpty() and
    self.ExpectedResult.Result.observable()
```
5. Unambiguity Outcome: The same precondition and operation should not yield conflicting expected results
```OCL
context Rule
inv DeterministicOutcome:
    Rule.allInstances()->forAll(r2 |
        if r2 <> self and r2.Precondition = self.Precondition and r2.Operation = self.Operation
        then r2.ExpectedResult.ResultStatus = self.ExpectedResult.ResultStatus
        else true
        endif)
```
\end{artifact}

\subsection{Prompt Used for compared E2E LLMs}
\label{sec:app_prompt_llm}

In the experiments, we compare our approach with end-to-end LLMs in generating test cases from regulatory documents. Representative models, including GPT-5, Grok-4, and DeepSeek-R1, are selected as baselines. Task-specific prompts are carefully designed for them to elicit the best performance in test case generation. The prompts used in Experiment I are presented below; for Experiment II, the composition is the same, except for the application domain and the running example.

\begin{markdowncode}
You are a Financial Requirement Modeling & Testing Expert, specializing in software requirements engineering, financial business rule analysis, and test case generation.
Your task is to generate test cases for the given financial-domain rules.

### Task Objective
Test Case Generation (only when the rule is testable)
* Based on the natural language rule, generate a set of high-quality test cases.
* Generation requirements:
  1. Content requirements
     * Each test case must be a JSON object (map).
     * Each test case should include multiple key elements, such as: actor, time, trading method, trading instrument, action, operation target, quantity, price, result, etc. (to be flexibly selected according to the rule content).
  2. Quality requirements
     * Test cases must cover both positive (success) and negative (failure) scenarios.
     * Test cases should reflect rule boundaries and constraint validation.
     * Test cases should not be duplicated and must be representative and distinguishable.
     * The output must be strictly in JSON array format.
* If the rule is not testable and test cases cannot be generated, do not generate any output.

### Input Format
Input consists of one natural language financial-domain rule.

### Output Format
Output test cases in JSON format:
[
  {
    "id": 1,
    "TradingInstrument": ...,
    "Time": ...,
    ...
  },
  {
    "id": 2,
    ...
  },
  ...
]

### Example
#### Input:
An investor who buys bond ETF shares through auction trading on the same day may sell them on the same day.
#### Output:
[
  {
    "id": 1,
    "Actor": "Investor",
    "Time": "Same day",
    "TradingMethod": "Auction trading",
    "Action": "Buy",
    "TradingInstrument": "Bond ETF shares",
    "Time2": "Same day",
    "Action2": "Sell",
    "Result": "Success"
  },
  {
    "id": 1,
    "Actor": "Investor",
    "Time": "Not the same day",
    "TradingMethod": "Auction trading",
    "Action": "Buy",
    "TradingInstrument": "Bond ETF shares",
    "Time2": "Same day",
    "Action2": "Sell",
    "Result": "Failure"
  },
  ...
]
\end{markdowncode}

\end{document}